\documentclass[secnumarabic,superscriptaddress,amssymb, nobibnotes, aps, twocolumn, prl]{revtex4-2}

\usepackage{amsmath}
\usepackage{mathrsfs}  % use to write fancy letters (e.g. for free energy)
\usepackage{mathtools} % use to write := im math
\usepackage{bbold}    % use to write fancy 1 for unity matrix
\usepackage{braket}    % lets you use the bra, ket notation
\usepackage{gensymb}   % for \degree

\usepackage{graphicx}      	% import figures
\usepackage{multirow}       % merge rows and columns in table
\usepackage{here} 			% display figures exactly where they are in code
\usepackage{url}		% display urls in a nice way
\usepackage[plainpages=false,pdfpagelabels=true, linkcolor=cyan,breaklinks,colorlinks=true]{hyperref}  % hyperrefs
\usepackage{soul}                   % allows to cross out text
\usepackage[dvipsnames]{xcolor}		% text in different colors
\usepackage{siunitx}
\usepackage{xspace}
\usepackage{booktabs}    % for prettier lines in table
\usepackage{ulem}

\newcommand{\cvs}{CsV$_3$Sb$_5$\xspace}
\newcommand{\rvs}{RbV$_3$Sb$_5$\xspace}
\newcommand{\Tc}{$T_{\text{c}}$\xspace}
\newcommand{\Tcdw}{$T_{\text{CDW}}$\xspace}

\begin{document}

\title{Out-of-equilibrium relaxation dynamics of the superconducting order parameter in \cvs}%

\author{Avi Shragai}%
\affiliation{Laboratory of Atomic and Solid State Physics, Cornell University, Ithaca, NY 14853, USA}

\author{Alexander Won}%
\affiliation{Laboratory of Atomic and Solid State Physics, Cornell University, Ithaca, NY 14853, USA}

\author{Jonathan M. DeStefano}%
\affiliation{Laboratory of Atomic and Solid State Physics, Cornell University, Ithaca, NY 14853, USA}

\author{Andrea Capa Salinas}
\affiliation{Materials Department, University of California Santa Barbara, Santa Barbara, CA, USA}
\author{Ganesh Pokharel}
\affiliation{Materials Department, University of California Santa Barbara, Santa Barbara, CA, USA}
\author{Stephen D. Wilson}
\affiliation{Materials Department, University of California Santa Barbara, Santa Barbara, CA, USA}

\author{B. J. Ramshaw}
\email{bradramshaw@cornell.edu}
\affiliation{Laboratory of Atomic and Solid State Physics, Cornell University, Ithaca, NY 14853, USA}
\affiliation{Canadian Institute for Advanced Research, Toronto, Ontario, Canada}

\date{\today}%

\begin{abstract}
	
The application of a time-varying strain field drives a superconducting order parameter out of equilibrium. How the order parameter relaxes back to equilibrium depends both on the structure of the superconducting gap and on the nature of quasiparticle scattering. We report the discovery of an ultrasonic attenuation peak inside the superconducting state of the kagome superconductor \cvs. This peak is the natural consequence of the order parameter relaxation time matching the ultrasonic drive frequency near \Tc. From the measured frequency dependence of the peak, we extract a microscopic scattering time of $\tau_N = 25$ ps. This timescale is two orders of magnitude longer than the elastic scattering time as determined by resistivity measurements, but is comparable to the inelastic scattering time determined by thermal transport. Within the conventional framework of order-parameter relaxation, this implies that elastic scattering is ineffective at relaxing the superconducting condensate, consistent with a sign-preserving $s$-wave state obeying Anderson's theorem.

%While sound attenuation peaks have been associated with exotic superconducting order parameters in the past, careful analysis shows that our data are consistent with a conventional, $s$-wave gap. \cvs stands apart from other BCS superconductors, which typically do not exhibit peaks in sound attenuation, because of its unusually large inelastic scattering rate. 

%Conventional BCS superconductors exhibit a well-understood suppression of ultrasonic attenuation as they are cooled below their critical temperature. Deviations from this behavior are rare, and features in the ultrasound attenuation beyond the BCS description are frequently a sign of unconventional superconductivity. Here, we present measurements of the ultrasound attenuation of kagome superconductors \cvs and \rvs through their superconducting transition temperatures, $T_c$. In \cvs, the attenuation of compressional sound exhibits a peak just below $T_c$, with temperature and frequency dependence consistent with a Landau-Khalatnikov model of order-parameter relaxation. Despite the seemingly exotic nature of the attenuation peak, careful analysis shows that our data are consistent with a conventional, $s$-wave superconducting state largely protected against non-magnetic disorder. The attenuation peak in \cvs does not on its own imply unconventional superconductivity but rather is most naturally interpreted as arising from order-parameter relaxation driven by an unusually large inelastic scattering rate.     

\end{abstract}

\maketitle

% \section{Introduction}

\begin{figure}[t]
    \centering
    \includegraphics[width=0.5\textwidth]{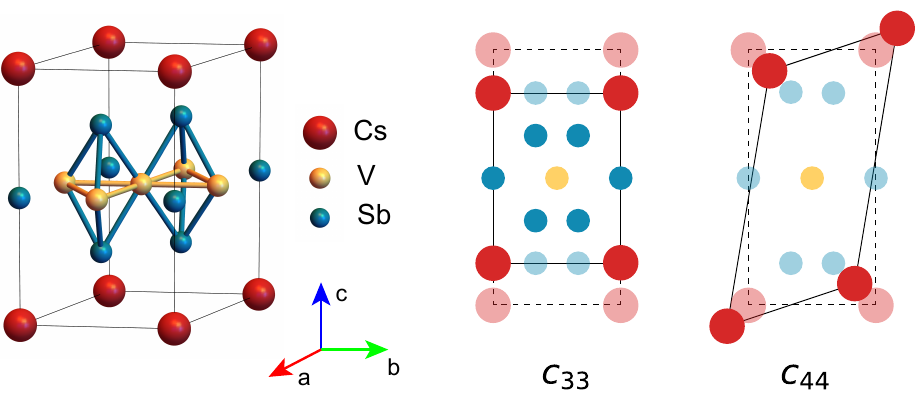}
    \caption{\textbf{Crystal structure and experimental strains}.
	Crystal structure of \cvs is shown on the left. In our experimental geometry, sound propagates along the crystal $c$ axis and probes the compressional modulus $c_{33}$ and the shear modulus $c_{44}$. 
	The atomic displacements associated with the $c_{33}$ and $c_{44}$ strains are shown on the right. 
	}
    \label{fig:modes}
\end{figure}

\textit{Introduction-} The AV$_3$Sb$_5$ family of kagome metals, with A= Cs, Rb, and K, exhibits rich phase diagrams that include both a charge density wave (CDW), with an onset temperature near \Tcdw$\approx 100$ K, and superconductivity (SC) below a critical temperature of \Tc$\approx1-3$ K \cite{ortizNewKagomePrototype2019,ortizCsV3Sb5Topological2020,jiangUnconventionalChiralCharge2021,yinSuperconductivityNormalStateProperties2021}. The cesium-based compound, \cvs, is particularly intriguing, with conflicting reports of time-reversal-symmetry breaking in both the charge-density-wave and superconducting states \cite{chaEvidenceTimereversalSymmetry2026,khasanovTimereversalSymmetryBroken2022,liNoObservationChiral2026,saykinHighresolutionPolarKerr2026}. Although some experiments favor a nodeless, anisotropic superconducting gap, the symmetry of the superconducting order parameter, and its relationship to the CDW state from which it emerges, remains unresolved \cite{mineObservationFermisurfacedependentAnisotropic2025,zhongNodelessElectronPairing2023,zhangNodelessSuperconductivityKagome2023,roppongiBulkEvidenceAnisotropic2023,muSWaveSuperconductivityKagome2021,duanNodelessSuperconductivityKagome2021, wilsonAV3Sb5KagomeSuperconductors2024,fernandesLoopcurrentOrderKagome2026,disanteKagomeMetals2026}.

A major impediment to identifying superconducting order parameters is the lack of phase-sensitive probes. With the exception of the landmark tri-crystal junction experiments in the high-$T_{\rm c}$ cuprates \cite{kirtleySymmetryOrderParameter1995}, the symmetry of a superconducting gap is usually inferred either from angle-resolved photoemission spectroscopy (ARPES), or by fitting power laws to quantities such as thermal conductivity or the microwave penetration depth. Both of these approaches, however, are difficult when the superconducting gaps are small, and neither can distinguish different fully-gapped states, such as $s-$wave, $s\pm$, $d_{x^2-y^2}\pm id_{xy}$, and $p_x\pm ip_y$---all of which have been proposed as candidates in AV$_3$Sb$_5$ \cite{guguchiaTunableUnconventionalKagome2023,leSuperconductingDiodeEffect2024,dengEvidenceTimereversalSymmetrybreaking2024,wangUnconventionalHysteresisDue2026,schultzSuperconductivityKagomeMetals2026,tazaiNematicChiralSuperconductivity2026}.

% One major impediment in our ability to identify superconducting order parameters is a distinct lack of phase-sensitive probes. With the exception of the landmark tri-crystal junction experiments in high-$T_{\rm c}$ cuprates, the symmetry of most superconducting gaps is inferred either from angle-resolved photoemission spectroscopy (ARPES), or by fitting power laws to quantities such as thermal conductivity or microwave penetration depth. Both of these approaches, however, are difficult when the superconducting gaps are small, and both are unable to distinguish different fully-gapped states, such as conventional $s-$wave, $s\pm$, and $d_{x^2-y^2}+id_{xy}$ \brad{cite crazy p-wave bullshit here}.

One way to distinguish between different fully-gapped states is by examining their response to quasiparticle scattering. In an $s-$wave state, elastic impurity scattering is unable to break Cooper pairs because the phase and amplitude of the gap is constant around the Fermi surface---a result known as Anderson's theorem \cite{andersonTheoryDirtySuperconductors1959}. For a gap with a winding phase or a sign change, however, elastic impurity scattering quickly suppresses \Tc \cite{abrikosovCONTRIBUTIONTHEORYSUPERCONDUCTING1960}. 

Experimentally, the sensitivity of the superconducting gap to elastic scattering is typically determined by introducing impurities and measuring the extent to which \Tc is reduced. A cleaner but less common method is to push the superconducting order parameter out of equilibrium and then measure how fast it relaxes back to equilibrium. The order parameter relaxation rate is determined both by the type of quasiparticle scattering and on the structure of the gap. An order parameter relaxation rate that is much slower than the elastic scattering rate---as determined, for example, from the low-temperature resistivity---suggests that elastic scattering does not relax the order parameter. This is most commonly interpreted as evidence for the applicability of Anderson's theorem---naively an indication of a fully gapped, $s$-wave state.

\begin{figure*}[t]
	\begin{center}
		\includegraphics[width=0.95\linewidth]{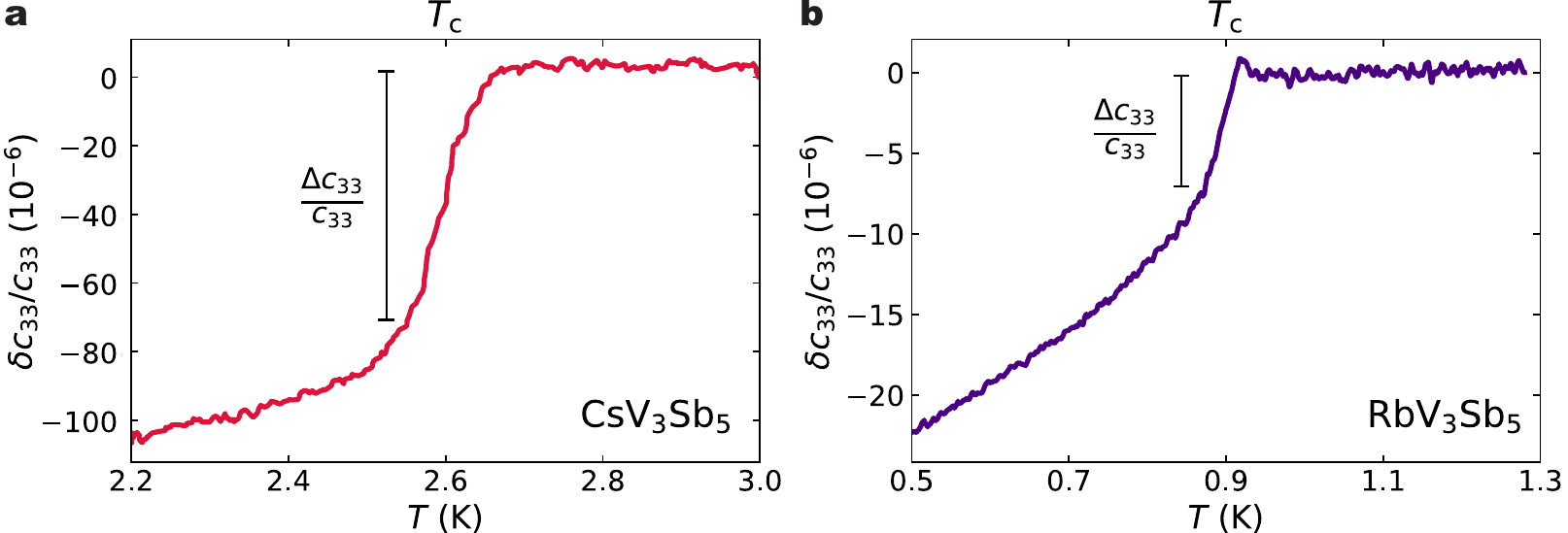}
	\end{center}
	\caption{ \textbf{Longitudinal elastic modulus jump at \Tc}.
	\textbf{a,} Fractional change in the longitudinal modulus $c_{33}$ of \cvs across its superconducting transition.
	\textbf{b,} Corresponding fractional change in $c_{33}$ of \rvs. The fractional discontinuity is significantly smaller and sharper in \rvs compared to \cvs. We attribute the rounding of the jump in \cvs to a combination of strain broadening of the superconducting \Tc, and relaxational effects on the sound velocity (Supplementary Figure 8 shows data from 0.46 GHz to 2.44 GHz).  
	}
	\label{fig:dcc-Tc}
\end{figure*}

% One way to test the sensitivity of the superconducting gap to disorder is to introduce impurities and examine how \Tc is reduced. A simpler---though rarely performed---method is to push the superconducing order parameter out of equilibrium and then to measure its relaxation rate. This order parameter relaxation rate will depend strongly on the type of quasiparticle scattering and on the structure of the gap, and is thus a test of whether a gap obeys Anderson's theorem.

Here we use sound attenuation---the rate at which sound energy dissipates in a crystal---to study the out-of-equilibrium dynamics of the order parameter in \cvs and \rvs. In the hydrodynamic limit (when the sound wavelength is long compared to the electronic mean free path), sound attenuation in a metal is produced when the sound wave distorts the Fermi surface and quasiparticles relax to the new equilibrium distribution. Sound attenuation is therefore governed by the availability of states at the Fermi level, and by the scattering mechanisms between those states \cite{khanSoundAttenuationElectrons1987a}. Conventionally, sound attenuation in a superconductor is exponentially suppressed when quasiparticles are gapped out below \Tc---a seminal prediction of BCS theory \cite{bardeenTheorySuperconductivity1957}. Deviations from this behavior are a signature of additional relaxation channels, such as order-parameter relaxation or collective modes \cite{monienUltrasoundAttenuationDue1987,bahlouliGapRelaxationIts1989}.

%Ultrasound attenuation in particular -- the rate at which sound energy dissipates in a crystal -- offers unique insight into the dissipative dynamics of the electronic system coupled to strain. Long-wavelength strain distorts the Fermi surface and generates heat as electrons scatter back to equilibrium. The attenuation, therefore, is governed by the availability of states at the Fermi level and their coupling to strain.\cite{khanSoundAttenuationElectrons1987a} When electronic excitations are gapped out in the superconducting state, the depletion of thermally excited quasiparticles results in a suppression of the attenuation on cooling below \Tc, a seminal prediction of BCS theory.\cite{bardeenTheorySuperconductivity1957} Deviations from this behavior are a signature of additional relaxation channels, such as absorption by collective modes or order-parameter relaxation. \cite{monienUltrasoundAttenuationDue1987,bahlouliGapRelaxationIts1989}   

We employ pulse-echo ultrasound to measure the ultrasonic attenuation of \cvs and \rvs through their SC transition temperatures. Contrary to the BCS prediction, we find a sharp peak in the attenuation of longitudinal sound in \cvs just below its superconducting \Tc. We obtain excellent agreement between the data and a simple model of order-parameter relaxation. The extracted timescale suggests that order-parameter relaxation in \cvs is controlled by a microscopic timescale that is two orders of magnitude longer than the elastic scattering time. We discuss possible origins of this scattering time and their implication for the symmetry of the superconducting gap. 

%Naively, this suggests that inelastic scattering dominates the order parameter relaxation in \cvs and that it is therefore a sign-preserving, $s$-wave superconductor. However, the situation may be more subtle in \cvs because, under specific assumptions about the order parameter and the location and strength of the scattering sites, kagome superconductors have been suggested to be more robust to elastic scattering \cite{holbaekUnconventionalSuperconductivityProtected2023}.

%Although LK attenuation peaks, which have been observed in heavy-fermion superconductors UBe$_{13}$, UPt$_3$, and UTe$_2$, are often associated with unconventional superconductivity, our data underscore that such a peak can also arise in an Anderson-protected superconductor when the inelastic scattering rate is sufficiently large.\cite{batlogg$ensuremathlambda$ShapedUltrasoundAttenuationPeak1985,bishopUltrasonicAttenuationUP$mathrmt_3$1984,kamatVanishingPhaseStiffness2026} Our result indicates that the superconducting state of \cvs is largely unaffected by non-magnetic impurity scattering, consistent with a sign-preserving $s$-wave gap.      

% \section{Methods}

\begin{figure*}[t]
	\begin{center}
		\includegraphics[width=0.95\linewidth]{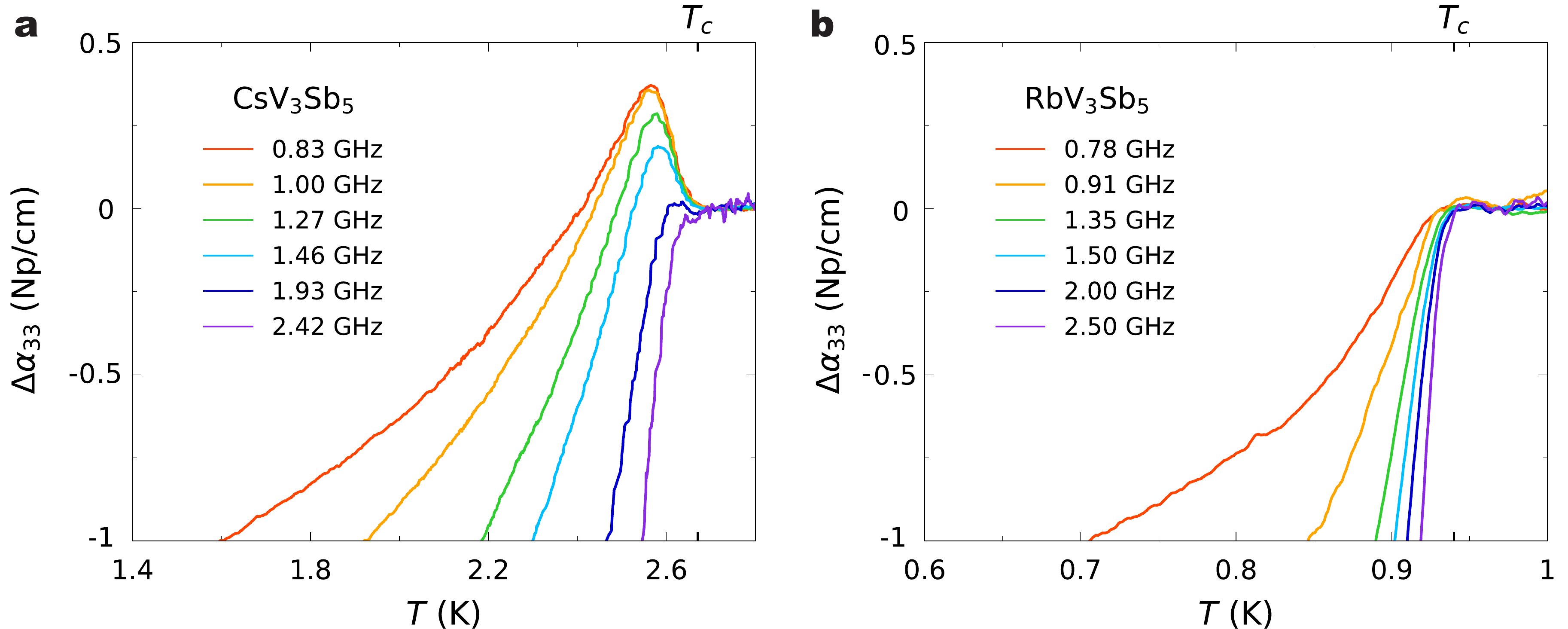}
	\end{center}
	\caption{\textbf{Longitudinal sound attenuation near \Tc}.
	\textbf{a,} Shows the temperature evolution of sound attenuation in \cvs at different ultrasonic frequencies. Sound attenuation is peaked just below \Tc at low frequencies---this peak decreases with increasing frequency and disappears entirely above $\sim$2 GHz.
	\textbf{b,} In contrast with \cvs, the sound attenuation in \rvs decreases monotonically below \Tc, conforming to the expectation for a BCS superconductor. 
	}
	\vspace{-0em}
	\label{fig:alpha-Tc}
\end{figure*}

\textit{Methods-} \cvs and \rvs single crystals were synthesized by the self-flux method. To perform pulse-echo ultrasound measurements, thin-film piezoelectric zinc oxide transducers were sputtered directly onto the exposed (001) sample surface \cite{theussSinglecomponentSuperconductivityUTe22024}. We sputter at an angle to produce mixed-mode transducers that apply both shear and longitudinal strain. These two modes travel at different speeds and separate in the time domain (an example of the raw pulse train is shown in Supplementary Figure 3). Longitudinal sound propagating along the crystallographic $c$ axis probes the $c_{33}$ compression modulus, and transverse sound probes the $c_{44}$ shear modulus. The crystal structure and relevant strain geometries are depicted in \autoref{fig:modes}. Further details of the sample synthesis and transducer fabrication are provided in the SI. 

% To perform pulse-echo ultrasound measurements, thin-film piezoelectric zinc oxide transducers were sputtered directly onto the sample surface. By controlling the deposition conditions, we selectively grew ZnO films with the $c$ axis either perpendicular or tilted relative to the sample surface normal, producing longitudinal or mixed-mode longitudinal and transverse transducers, respectively \cite{theussSinglecomponentSuperconductivityUTe22024}. Further details of the transducer fabrication process are provided in the SI. For the exposed (001) faces used in this study, sound propagated along the crystallographic $c$ axis and our measurements were sensitive to the compressional elastic modulus $c_{33}$ and the shear modulus $c_{44}$. The crystal structure and relevant strain geometries are depicted in \autoref{fig:modes}.

Changes in the sound velocity and attenuation were measured with a conventional pulse-echo technique. The ZnO transducer is driven with an RF voltage pulse exciting a strain wave that propagates through the sample. After reflecting from the opposing face of the crystal the strain pulse generates a voltage when it is incident on the transducer, which is recorded on an oscilloscope. The phase and amplitude of the reflected pulses are used to extract changes in the sound velocity and attenuation as a function of temperature \cite{PhysicalAcousticsSolid2005}. Further details are provided in the SI.

\textit{Results-} The changes in $c_{33}$ of \cvs and \rvs near their superconducting transitions are plotted in \autoref{fig:dcc-Tc}. Discontinuities of approximately 60 parts per million and 10 ppm occur in $c_{33}$ at the \Tc's of \cvs and \rvs, respectively. These discontinuities are expected on general thermodynamic grounds---they are the longitudinal strain equivalent of the jump up in specific heat at a second order phase transition, and can be quantitatively related to that jump through an Ehrenfest relation \cite{ghoshThermodynamicEvidenceTwocomponent2021,theussSinglecomponentSuperconductivityUTe22024}. The magnitudes of the discontinuities are consistent with the strong pressure dependence of the superconducting \Tc's in both compounds \cite{chenDoubleSuperconductingDome2021,yuUnusualCompetitionSuperconductivity2021,liDiscoveryConjoinedCharge2022}.

% The magnitude of the discontinuities is comparable to that observed in other superconducting materials \brad{cite}, indicating sizeable coupling between the superconducting order parameter and strain in both \cvs and \rvs. This observation is consistent with the strong pressure dependence of the superconducting transition temperatures in both compounds.\cite{chenDoubleSuperconductingDome2021,yuUnusualCompetitionSuperconductivity2021,liDiscoveryConjoinedCharge2022}

Having established that longitudinal $c$-axis strain couples to superconductivity in \cvs and \rvs, we turn to the ultrasonic attenuation, shown in \autoref{fig:alpha-Tc}a. In \cvs, the ultrasonic attenuation as a function of temperature increases suddenly immediately below the superconducting \Tc before decreasing on further cooling. The magnitude and width of the attenuation peak decrease as the ultrasound frequency is increased---above approximately 2 GHz, no anomaly is visible in the data. This behavior was reproduced in a second sample shown in the SI. 

In contrast, \rvs shows no sign of a peak in sound attenuation over a similar range of frequencies (\autoref{fig:alpha-Tc}b). Instead, the attenuation of \rvs decreases monotonically on cooling into the superconducting state---more conventional behavior for a superconductor. 
% The contrast between the behavior of \cvs and \rvs, despite their similar crystal structures and high-temperature CDW transitions, constrains possible mechanisms for the attenuation peak.  

% \section{Modeling the attenuation peak}

\textit{Modeling the attenuation peak-} The observation of an ultrasonic attenuation peak in the superconducting state of \cvs is unusual \cite{phillipsUltrasonicAttenuation5001969, claiborneStudyAttenuationUltrasonic1964,trivisonnoUltrasonicAttenuationLongitudinal1971} but not unheard of. Several heavy-fermion systems, including UTe$_2$, UBe$_{13}$, and UPt$_3$, exhibit peaks in longitudinal sound attenuation just below their \Tc's, similar to what we find in \cvs \cite{batlogg$ensuremathlambda$ShapedUltrasoundAttenuationPeak1985,bishopUltrasonicAttenuationUP$mathrmt_3$1984,kamatVanishingPhaseStiffness2026}. % While a number of intrinsic mechanisms for the attenuation peaks in heavy fermions have been proposed---including pair breaking and absorption by collective modes \cite{goldingObservationCollectiveMode1985}---
Miyake and Varma demonstrated that the peaks in UBe$_{13}$ and UPt$_3$ were consistent with a Landau-Khalatnikov (LK) model order-parameter relaxation, where the relaxation timescale crosses the ultrasound period near \Tc \cite{miyakeLandauKhalatnikovDampingUltrasound1986}. Recent ultrasound measurements of UTe$_2$ agree exceptionally well  with this model \cite{kamatVanishingPhaseStiffness2026}. We therefore test whether an LK model can explain the attenuation peak of \cvs.

% The observation of an ultrasound attenuation peak in the superconducting state of \cvs is unusual---sound attenuation in most superconductors decreases monotonically below \Tc. This prediction of BCS theory has been verified in numerous materials and our measurements of \rvs are naturally interpreted in this way \cite{phillipsUltrasonicAttenuation5001969, claiborneStudyAttenuationUltrasonic1964,trivisonnoUltrasonicAttenuationLongitudinal1971}.

%Instead, when a metal undergoes a superconducting transition, BCS theory predicts that the attenuation will be exponentially suppressed as the superconducting gap grows. \cite{bardeenTheorySuperconductivity1957,kadanoffUltrasonicAttenuationSuperconductors1966} This prediction has been verified in numerous materials and our measurements of \rvs are naturally interpreted in this way. \cite{phillipsUltrasonicAttenuation5001969, claiborneStudyAttenuationUltrasonic1964,trivisonnoUltrasonicAttenuationLongitudinal1971} 

The LK mechanism considers the effect of order-parameter relaxation on ultrasonic attenuation near the critical temperature of a continuous phase transition \cite{landauCollectedPapersLD1965}. An order parameter driven out of equilibrium by an applied strain field relaxes back to equilibrium with a timescale $\tau_\Delta$, where $\Delta$ emphasizes that this is the timescale of macroscopic gap dynamics, and not the timescale of microscopic quasiparticle relaxation. When $\tau_\Delta$ is far removed from the inverse angular frequency of the applied strain, $1/\omega$, the gap magnitude is either always nearly in equilibrium ($\omega \tau_\Delta\ll 1$) or acts as a quasi-static background for the strain ($\omega \tau_\Delta \gg 1$). If the two timescales are comparable, however, energy from the sound wave is lost as it generates entropy in the superconductor \cite{miyakeLandauKhalatnikovDampingUltrasound1986,bahlouliGapRelaxationIts1989}. 

Attenuation due to the LK mechanism takes the form
\begin{equation}
\label{eqn:LK-damping}
\alpha_{LK}(\omega, T) = \frac{2}{v}\left|\frac{\Delta v}{v}\right|\frac{\omega^2\tau_\Delta(T)}{1 + \left(\omega\tau_\Delta(T)\right)^2},
\end{equation}
where $v$ is the speed of sound and $\Delta v/v = (\Delta c/c)/2$ is the fractional jump in the sound velocity at \Tc---a quantity we measure (\autoref{fig:dcc-Tc}). A full derivation of \autoref{eqn:LK-damping} is provided in the SI. The temperature dependence comes entirely from $\tau_{\Delta}(T)$, which diverges at the phase transition. From this expression, it is clear that a necessary condition for a peak in the sound attenuation is $\omega \tau_\Delta(T) \approx 1$. 

To model the data, we also include a conventional BCS-like term that captures the decrease in sound attenuation due to the opening of the superconducting gap \cite{bardeenTheorySuperconductivity1957}:
\begin{equation}
\label{eqn:alpha-BCS}
\alpha_{SC}(\omega, T) = \frac{2 A_n \omega^2}{1 + e^{\Delta(T)/k_B T}},
\end{equation}
where $A_n\omega^2$ captures the normal state attenuation, and the temperature dependence of the gap is approximated by $\Delta(T) = \Delta_0\tanh(1.74\sqrt{T_{\text{c}}/T - 1})$. Note that, while we fit the entire temperature dependence of the sound attenuation, we are largely interested in the behavior of the peak near \Tc, and therefore details of how we model $\Delta(T)$ (i.e. whether we use one or two gaps) are not crucial for our analysis.

\begin{figure*}[t]
	\begin{center}
		\includegraphics[width=0.95\linewidth]{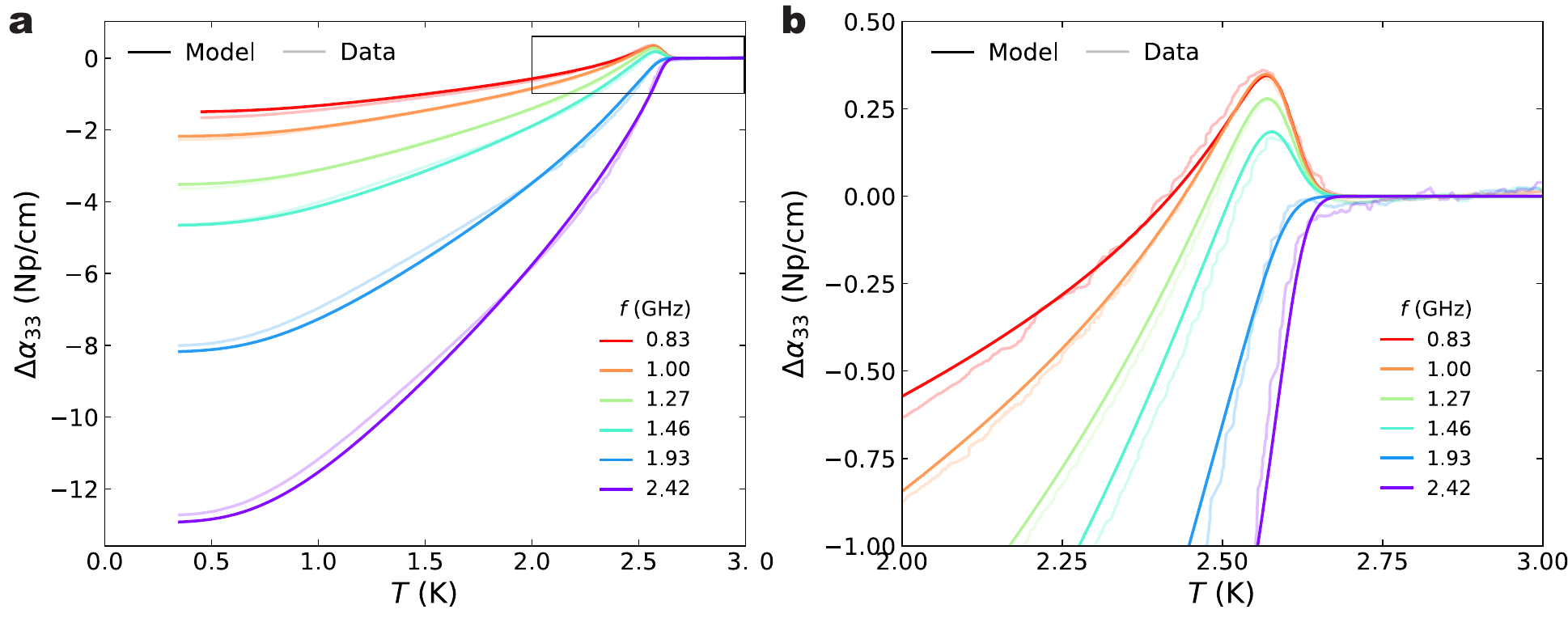}
	\end{center}
	\caption{ \textbf{Fit to the data with the LK model}.
	Fit to the data using the sum of order-parameter-relaxation and BCS quasiparticle contributions. Opaque traces are the model, and semitransparent traces are the data. \textbf{a,} Shows the full temperature range and \textbf{b,} shows a narrower range of temperatures near \Tc, as indicated by the bounding box in \textbf{a}. From the LK model, we extract a characteristic timescale for the order-parameter relaxation to be $\tau_N = 25$ ps. 
	}
	\label{fig:alpha-fit}
\end{figure*}

\autoref{fig:alpha-fit} shows a fit to the \cvs data using the sum of \autoref{eqn:LK-damping} and \autoref{eqn:alpha-BCS}. All six frequencies are fit simultaneously using the same set of model parameters. We find excellent agreement with the data using a temperature-dependent relaxation time of the form
\begin{equation}
\label{eqn:tau-T}
\tau_\Delta(T) = \frac{\tau_N}{\sqrt{1 - T/T_{\text{c}}}},
\end{equation}
with an extracted characteristic timescale of $\tau_N = 25$ ps (we discuss our choice of \autoref{eqn:tau-T} in more detail below). A detailed discussion of the remaining fit parameters is presented in the SI.

As a check of the applicability of the LK model, we also measure the attenuation of shear sound along the $c$ axis of \cvs (see SI). We observe no attenuation peak for shear sound, consistent with the requirement that shear strain couples quadratically to the order parameter (as opposed to the linear-in-strain coupling for longitudinal strain, which is required for an LK peak).

The absence of a peak in \rvs is also explained within the LK framework: we measure the pre-factor of \autoref{eqn:LK-damping}, $2\Delta v/v^2$, to be more than an order of magnitude smaller for \rvs than for \cvs. This is sufficient to completely suppress the attenuation peak (i.e. to make the rise in attenuation due to the peak smaller than the drop due to the opening of the gap, see SI for details).

% \brad{I greatly reduced the ``it's surprising because it gets smaller at high frequencies'' paragraph because it's only surprising to us. Maybe we can add in a sentence about ``note the opposite dependence compared to'' somewhere else, but I think this is answering a question no one is asking (except us). }
%The appearance of an LK peak in \cvs is nevertheless surprising in two ways. First, the frequency dependence of the LK attenuation alone in \autoref{eqn:LK-damping} naively suggests that the magnitude of the attenuation peak should grow with increasing frequency -- the opposite trend we observe in our data. The seemingly uncharacteristic behavior we observe in \cvs is due to the competition between the LK peak, which increases $\sim \omega$ and is shifted away from \Tc at higher frequency, and the reduction of attenuation due to the opening of the gap, which increases $\sim \omega^2$. At sufficiently high frequency, then, the peak is swallowed by the BCS suppression of attenuation.

% \section{Analysis of the relaxation timescale}

\textit{Analysis of the relaxation timescale-} Having established that the temperature and frequency dependence of the attenuation peak in \cvs are well-described by an LK model, we now ask what the fitted relaxation time tells us about the superconducting order parameter. 

The character of gap relaxation depends on the dimensionless product of the superconducting gap and the scattering time relevant to gap relaxation: $\Delta(T)\tau_N/\hbar$, where $\tau_N$ is the microscopic scattering time of individual quasiparticles \cite{miyakeLandauKhalatnikovDampingUltrasound1986}. Note that the timescale $\tau_N$ remains finite at \Tc and is distinct from $\tau_\Delta$, which depends on the gap susceptibility and therefore diverges at \Tc. 

In the regime $\Delta(T)\tau_N/\hbar \ll 1$, many microscopic scattering events occur during the pairing timescale $\hbar/\Delta$, and the return to equilibrium is limited by the order-parameter susceptibility. This gives a characteristic timescale near the critical temperature proportional to the thermal timescale $\hbar/k_BT_{\text{c}}$ multiplied by the gap susceptibility, which diverges as $\sim (1 - T/T_{\text{c}})^{-1}$ \cite{miyakeLandauKhalatnikovDampingUltrasound1986}.

In the opposite limit, when $\Delta(T)\tau_N/\hbar \gg 1$, the microscopic scattering time, $\tau_N$, is the bottleneck and controls the gap relaxation. Because only a fraction---of order $\Delta(T)/k_B T_{\text{c}}$---of thermally excited states participate in gap relaxation, the collective relaxation time exceeds $\tau_N$ by a factor of $k_B T_{\text{c}}/\Delta(T) \sim \sqrt{1 - T/T_{\text{c}}}$ \cite{tinkhamIntroductionSuperconductivity2004}.

Detailed calculations, given by Miyake and Varma and by Schmid, support this intuitive picture and show that
\begin{equation}
\label{eqn:relaxation-cases}
\tau_\Delta(T) \sim \begin{cases}
\frac{\hbar}{k_BT_{\text{c}}}\frac{1}{1 - T/T_{\text{c}}} & \Delta(T)\tau_N/\hbar \ll 1 \\
\frac{\tau_N}{\sqrt{1 - T/T_{\text{c}}}} & \Delta(T)\tau_N/\hbar \gg 1,
\end{cases}
\end{equation}
with order-one numerical prefactors that depend on the momentum-dependence of the gap and on the Fermi surface geometry \cite{miyakeLandauKhalatnikovDampingUltrasound1986,schmidApproachEquilibriumPure1968}. \autoref{eqn:relaxation-cases} is applicable as long as the superconducting transition is mean-field-like: the sharp specific heat jump (and sharp elastic modulus drop, \autoref{fig:dcc-Tc}) justifies this for the AV$_3$Sb$_5$ family.

As shown in \autoref{fig:alpha-fit}, our data are consistent with $\tau_\Delta \sim \sqrt{1 - T/T_{\text{c}}}$, and attempts to fit with $(1 - T/T_{\text{c}})^{-1}$ produce large quantitative and qualitative discrepancies (see SI). This indicates that the attenuation peak in \cvs occurs in the regime where $\Delta(T)\tau_N/\hbar \gg 1$, and that gap relaxation is limited by the quasiparticle scattering time rather than by the gap susceptibility. The remaining question is what microscopic process is associated with the extracted $\tau_N = 25$ ps.

We first consider elastic quasiparticle scattering. We estimate the elastic scattering time, $\tau_{el}$, near \Tc using resistivity data we obtained on the same sample (see SI) and fermi surface parameters (i.e. masses and carrier densities) from the literature \cite{chapaiMagneticBreakdownTopology2023,shresthaNontrivialFermiSurface2022,guoDistinctSwitchingChiral2024,ganMagnetoSeebeckEffectAmbipolar2021}. We obtain $\tau_{el} = m^*/ne^2\rho_0 \approx 0.6$ ps. This is nearly two orders of magnitude shorter than what we extract from the LK peak and would place us in the $\Delta(T)\tau_{el}/\hbar\ll 1$ regime---inconsistent with the data. We therefore turn to inelastic processes as a possible driver of gap relaxation. 

%It is worth emphasizing that if \rvs had a time-reversal-symmetry breaking gap, such as $p_x\pm ip_y$, then the relaxation rate would be set by the elastic scattering and we would observe an LK peak in that material as well---the fact that we do not observe one suggest that OP relaxation in \rvs is also not governed by elastic scattering. 

% The inelastic scattering time, $\tau_{in}$, can be estimated from both the electrical resistivity and from the electronic Lorenz number extracted from thermal transport. Both estimates yield an inelastic timescale on the order of 10 ps near \Tc. Details of these estimates are presented in the SI.

The inelastic scattering time, $\tau_{in}$, can be estimated from the electronic Lorenz number extracted from thermal transport. This yields an estimate of $\tau_{in} \approx 26$ ps near \Tc \cite{menilElectronelectronElectronphononCollision2026}. Details of this estimate is presented in the SI. This timescale---though unusually short for a good metal at these temperatures---is consistent with the value we obtain from our LK fit, supporting inelastic scattering as the simplest explanation for the order parameter relaxation peak we find in \cvs. To our knowledge, \cvs is the first non-heavy-fermion superconductor in which an ultrasound attenuation peak has been quantitatively described by Landau-Khalatnikov order-parameter relaxation.

\textit{Discussion-} Traditionally, the appearance of an order parameter relaxation peak in ultrasonic attenuation has been taken as an indication of unconventional superconductivity. This is because an unconventional superconducting gap is sensitive to fast elastic impurity scattering, placing the system in the regime where the OP relaxation time is limited by $\hbar /k_B T_{\text{c}}$. This timescale---typically of order a few picoseconds---is much smaller than the hundreds-of-picoseconds ultrasound period, resulting in an LK peak as the OP relaxation time diverges and the condition $\omega \tau_\Delta = 1$ is satisfied near \Tc. This situation---where elastic scattering dominates and the relaxation time is limited by the gap susceptibility---is realized in the unconventional superconductor UTe$_2$ \cite{kamatVanishingPhaseStiffness2026}.

In $s$-wave superconductors, on the other hand, OP relaxation is governed by the inelastic scattering time at \Tc---typically several nanoseconds or longer for \Tc's of a few kelvin---and the condition $\omega \tau_\Delta = 1$ is never satisfied for experimentally accessible ultrasound frequencies\footnote{One can in principle perform ultrasound measurements to essentially arbitrarily low frequencies. The issue is that the LK peak amplitude increases linearly with frequency. Thus while it might be possible to meet the $\omega \tau_\Delta = 1$ condition in, for example, aluminum or lead using MHz-frequency ultrasound, the LK peak would be 1000 times smaller than what we observe in our experiments in the GHz range and would therefore disappear into the noise.}, which is why LK peaks have never before been observed in conventional superconductors \cite{mitchellElectronphononScatteringSilver1985,doezemaMagneticSurfaceLevels1972,wegehauptMeasurementAnisotropicElectronphonon1978}. 

Our observation of an LK peak in \cvs is therefore unusual: the characteristic timescale---$\tau_N = 25$ ps---is much longer than the elastic scattering time inferred from resistivity, but it is also much shorter than the inelastic scattering time at \Tc in a conventional superconductor such as aluminum. There are two possible explanations: 1) either \cvs has a conventional $s$-wave gap but has exceptionally strong inelastic scattering; or 2) \cvs has an unconventional gap but elastic impurity scattering is unusually inefficient at relaxing the order parameter. 

The first scenario is the most straightforward interpretation of our data and is supported by the independently estimated inelastic scattering time extracted from thermal transport and resistivity: it appears that \cvs does indeed have exceptionally strong inelastic scattering. The second case is consistent with recent theoretical work \cite{holbaekUnconventionalSuperconductivityProtected2023}, which shows that, under specific assumptions regarding the superconducting order parameter and scattering potential, certain unconventional superconducting states on the kagome lattice can exhibit an anomalously weak sensitivity to elastic scattering. While this scenario provides a possible route to compatibility between our observation and an unconventional order parameter, it remains to be determined whether the required form of disorder protection is realized in \cvs.

\newpage

\section{Data Availability}

Data measured in this work are freely available at \url{http://github.com/CHiLL-Ramshaw/manuscripts-supporting_data/tree/main/2026_CsV3Sb5_LK}

\section{Acknowledgments}
B.J.R. and A.S. acknowledge helpful discussions with Mark Fischer and Andreas Kreisel. Research at Cornell was supported by the Department of Energy, Office of Basic Energy Sciences Award No. DE-SC-0026003 (ultrasound measurements and data analysis) and made use of facilities supported by the Cornell Center for Materials Research (transport measurements and sample characterization). S.D.W. and A.C.S. acknowledge support via the UC Santa Barbara NSF Quantum Foundry funded via the Q-AMASE-i program under award DMR-1906325 (sample growth and characterization). J.M.D. acknowledges support from the Klarman fellowship and the Kavli Institute at Cornell experimental postdoctoral fellowship (transport measurements).

\end{document}